%% file: aisafety4se.tex
\pdfoutput=1

\documentclass[sigconf,anonymous=false,review=false,authorversion=true, screen=true, nonacm=true, authordraft=false, balance=false]{acmart}

\makeatletter
\def\@ACM@checkaffil{
    \if@ACM@instpresent\else
    \ClassWarningNoLine{\@classname}{No institution present for an affiliation}%
    \fi
    \if@ACM@citypresent\else
    \ClassWarningNoLine{\@classname}{No city present for an affiliation}%
    \fi
    \if@ACM@countrypresent\else
        \ClassWarningNoLine{\@classname}{No country present for an affiliation}%
    \fi
}
\makeatother

\AtBeginDocument{%
  \providecommand\BibTeX{{%
    \normalfont B\kern-0.5em{\scshape i\kern-0.25em b}\kern-0.8em\TeX}}}

\acmConference[Conference acronym 'XX]{Make sure to enter the correct
  conference title from your rights confirmation emai}{June 03--05,
  2018}{Woodstock, NY}
\acmPrice{15.00}
\acmISBN{978-1-4503-XXXX-X/18/06}

\usepackage{etoolbox,xspace}
\usepackage{footmisc}
\usepackage{tablefootnote}
\usepackage{etoolbox}

\usepackage{fancyhdr}

\begin{document}

\title{AI Safety Subproblems for Software Engineering Researchers}

\author{David Gros}
\email{dgros@ucdavis.edu}
\affiliation{%
  \institution{University of California, Davis}
}

\author{Prem Devanbu}
\email{ptdevanbu@ucdavis.edu}
\affiliation{%
  \institution{University of California, Davis}
}

\author{Zhou Yu}
\email{zy2461@columbia.edu}
\affiliation{%
  \institution{Columbia University}
}

\renewcommand{\shortauthors}{Gros et al.}

\begin{abstract}
In this 4-page 
manuscript we discuss the problem of long-term AI Safety from a Software Engineering (SE) research viewpoint. We briefly summarize long-term AI Safety, and the challenge of avoiding harms from AI as systems meet or exceed human capabilities, including software engineering capabilities (and approach AGI / ``HLMI'')
. We perform a quantified literature review suggesting that AI Safety discussions are not common 
at SE venues. We make conjectures about how software might change with rising capabilities, and categorize ``subproblems'' which fit into traditional SE areas, proposing how work on similar problems might improve the future of AI and SE.
\end{abstract}

\ccsdesc[500]{Software and its engineering}
\ccsdesc[500]{Computing methodologies~Artificial intelligence}
\ccsdesc[200]{Social and professional topics~Computing / technology policy}

\keywords{AI safety, AGI, Transformative AI, AI4SE, Code Generation}



\newcommand\redtodo[1]{\textcolor{red}{#1}}
\input{cite_vars}

\newcommand{\myaispara}[1]{\vspace{1mm} \noindent{\underline {\em #1}\,}}


\maketitle

\section{Introduction}


The rise of data-driven techniques has led to rapid progress in Artificial Intelligence (AI). Since 2010, compute 
power used to train the largest AI models has doubled approximately every 6 months \cite{sevilla2022compute}. The total private dollar investment in AI is more than 10 times what it was a decade ago \cite{maslej2023ai}.
Software Engineering (SE) researchers are increasingly taking note and contributing to this advance. When examining papers at large SE venues\footnote{FSE, ICSE, ASE, ISSTA\label{se_footnote}} in 2022, we estimate \SePercentAiNew mentioned AI/ML/DL terms in their title or abstract\footnote{based on a keyword match in Semantic Scholar corpus. \href{https://github.com/DNGros/aisse/blob/main/ai_terms_regex.txt}{\url{https://github.com/DNGros/aisse/blob/main/ai_terms_regex.txt}}}. This is in comparison to \SePercentAiOld in 2012.

The rise of AI for Software Engineering (AI4SE) and Software Engineering for AI (SE4AI) requires making \emph{AI Safety} a concern of SE researchers. AI Safety refers to avoiding harm from AI. We endorse arguments that the most pressing consideration here is safety at the limit, \emph{i.e.} safety as AI begins to match or exceed human capabilities in all domains. 
The collection of such systems is referred to as High Level Machine Intelligence (HLMI) \cite{GraceSDZE17}, and is similar to concepts like Artifical General Intelligence (AGI) or Transformative AI (TAI).
Failure to make progress on HLMI safety can have catastrophic consequences, including the extinction of humanity. SE researchers have a role to play in either exacerbating the dangers, or helping reduce risk. We will discuss why, and provide helpful conceptions of concrete AI problems, that resemble existing SE problems. 

We believe that safety implications of highly advanced AI are infrequently discussed at SE conferences, despite the increase in AI4SE work. 
We explore this hypothesis through a citation graph analysis\footnote{We use the citation graph (rather than traditional keyword search like \cite{DBLP:journals/corr/abs-2002-05671}) to hopefully better approximate works that give any consideration to long-term AI safety, even if it was not enough of a focus to appear in searchable fields.}.
We identify \Numfwais works that are ``foundational'' in HLMI safety (referred to as \fais)\footnote{\href{https://github.com/DNGros/aisse/blob/main/foundation_papers.csv}{\url{https://github.com/DNGros/aisse/blob/main/foundation_papers.csv}}}. 
The selection of the \fais works is informed by querying all references in a set of survey papers (\autoref{secextrarescources}) and examining works that are in multiple surveys. We consider a citation to any of these \Numfwais~works to be a proxy signal that a paper \textit{considers} concepts of long-term AI Safety (referred to as \cais). We use the Semantic Scholar API and corpus \cite{lo-etal-2020-s2orc} to query citations to the \fais. The earliest paper included in \fais is from 1960 \cite{Wiener1960SomeMA}. We filter \cais to be from 2012 to March 2023.

We identify a total of \FwaisTotalCitesDedup~unique \cais works making \FwaisTotalCitesDuplicates~citations to the \fais works. 
\autoref{tab:cais_table} shows counts by venue. \Numcaiscsfield \cais have ``Computer Science'' as an identified field.

Overall, \cais papers are relatively rare at major computing conferences\footnote{using a conference list from \href{http://csconferences.org/}{csconferences.org}}. Unsurprisingly, such topics are discussed most often in top ML-focused conferences (NeurIPS, ICLR, ICML, KDD) with \PcMlCais papers, or at AI-specific conferences (AAAI, IJCAI) with \PcAiCais papers\footnote{This count includes only the main technical conference, but misses some safety related workshops like SafeAI@AAAI and AISafety@IJCAI}. Notably, only \PcSeCais papers in SE conference venues\footref{se_footnote} reference \fais work out of \PcSeTotal papers in the corpus at those venues\footnote{Includes \cite{Tambon2021HowTC} which the corpus labels as in Automated Software Engineering  conference when actually in Automated Software Engineering journal}. 
Just \PcSejournalsCais other \cais papers appear in major SE journals out of \PcSejournalsTotal entries. 
About half of the \cais works are arXiv preprints or uncategorized. Scientific work often never reaches a publication, but AI Safety culturally relies heavily on self-publishing or web forums.

This analysis has many limitations. 
Large amounts of research (like in AI robustness or interpretability) is safety-motivated, even if it never references the \fais discussions of long-term AI. 
So while this analysis only approximates existing consideration of HLMI Safety, it suggests there is currently likely limited sharing of ideas between the SE and AI Safety communities. In this manuscript we hope to help bridge some of the gap.



\vspace{-0.2em}
\begin{table}[htbp]
    \centering
    \fontsize{7}{9}\selectfont
    \begin{tabular}{p{0.45\linewidth} rr}
        \hline
        Domain & \parbox[t]{0.1\linewidth}{\centering \cais} & \parbox[t]{0.1\linewidth}{\centering All}\\
        \hline
        \textbf{SE \hfill \fontsize{5}{7}\selectfont (ICSE, FSE, ASE, ISSTA)} & \PcSeCais & \PcSeTotal \\
        ML \hfill \fontsize{5}{7}\selectfont (ICLR, ICML, NeurIPS, KDD) & \PcMlCais & \PcMlTotal \\
        AI \hfill \fontsize{5}{7}\selectfont (AAAI, IJCAI) & \PcAiCais & \PcAiTotal \\
        NLP \hfill \fontsize{5}{7}\selectfont (ACL, EMNLP, NAACL) & \PcNlpCais & \PcNlpTotal \\
        Computer Vision \hfill \fontsize{5}{7}\selectfont (CVPR, ECCV, ICCV) & \PcCvCais & \PcCvTotal \\
        PL \hfill \fontsize{5}{7}\selectfont (PLDI, POPL, ICFP, OOPSLA) & \PcPlCais & \PcPlTotal \\
        Other Conference or Workshop & \PcOtherconfCais & \\
        \hline
        SE Journal \hfill \fontsize{5}{7}\selectfont (TSE, JSS, ESE, IST, TOSEM) & \PcSejournalsCais & \PcSejournalsTotal \\
        Other Journal or Workshop\tablefootnote{Some workshops appear classified as journals in the corpus} & \PcOtherjournalCais & \\
        \hline
        arXiv CS (and not other) & \PcArxivCais & \PcArxivTotal \\
        Venue Unknown / Other arXiv / Other & \OtherCais &  \\
        \hline
    \end{tabular}
    \caption{Estimated counts of works considering HLMI safety in major SE conferences and a subset of other fields.}
    \label{tab:cais_table}
    \vspace{-3.5em}
\end{table}
\vspace{-3.5em}

\clearpage

\section{AI Safety in 1-page}

This page concisely summarizes AI Safety concepts as background.

\myaispara{AI Safety Timelines:} A key consideration in safety is the timeline for how and when we reach systems more capable than humans in key parts of the economy, and in advancing science. There is evidence that AI researchers expect this could happen soon. A 2016 survey from \citet{GraceSDZE17} and a 2022 followup \cite{aiimpacts2022} asked \char`\~700 authors who published at ML/AI conferences to predict the arrival of HLMI, defined as ``when unaided machines can accomplish every task better and more cheaply than human workers''. The aggregate response put a 50\% chance by 2059. 

\citet{roser_2023} discusses this survey, three other independent surveys of experts and non-experts, and models based on estimates of computation needed for intelligence tasks. This and other surveys \cite{epoch2023literaturereviewoftransformativeartificialintelligencetimelines} suggest significant (>50\%) likelihood of highly capable AI in this century (thus relevant for people born today), and appreciable likelihood (>20\%) within two decades. Given the transformative effect of this event, probabilities on this scale must be taken seriously. 

\emph{This is the time} to attend to AI safety.

\myaispara{Takeoff Speed} "Takeoff" estimates how quickly AI could progress from near-human capabilities to vastly more capable than human (superintelligence, or ASI). More formal definitions try to define this in terms like changes to world economic output \cite{sidewaysviewTakeoffSpeeds}.

It seems probable the physical bounds for intelligence are vastly higher than what the human brain performs (which runs on about 20 watts 
and has less mass than some laptops). Because HLMI could do tasks like automating AI research and hardware development, recursive self improvement could lead to unexpected progress, creating a ``singularity'' in development. 
Takeoff speed is uncertain, but might mean we must ``get AI safety right the first time''.

\myaispara{The Alignment Problem} Alignment refers to making a system follow designers' intended goal. 
An AI system may prefer states about the world that it optimizes toward (\emph{eg}, a chess AI prefers states of the world where it wins). Aligned HLMI should prefer states that humanity also prefers. There are nuances to alignment, such as outer- and  inner-alignment, and objective robustness \cite{hubinger2019risks, shah2022goal, DBLP:journals/corr/abs-2105-14111}.

Three examples of real-world AI alignment failures include \textbf{(1)} the Bing chatbot doing a web search for a user's name, and then threatening them \cite{bing_chat_lw}, \textbf{(2)} Social media AI algorithms optimized for clicks/engagement time when this objective misaligned with societal benefit \cite{Saurwein2021AutomatedTT, bergen2019youtube, Stray2020AligningAO}, \textbf{(3)} Code generators that produce buggy code when they are capable of producing correct code \cite{codex, Jesse2023Simple}.


\myaispara{Threats from Misaligned HLMI} Intelligence is an ability to achieve goals in a wide range of environments \cite{legg2007collection}. ``\textit{Instrument convergence}’’ suggests many sub-goals are useful for all intelligent agents (like sub-goals of having access to resources or ensuring self preservation). Unless well-aligned, HLMI systems might cause catastrophic or existential harm to humanity while achieving goals. This is analogous to how humans might chop a forest or use pesticides not out of hate for animals living there, but to achieve other goals. Systems need not be malicious to cause harm, only very capable. \cite{bostrom2017superintelligence}


\myaispara{Current Agendas} Work is often categorized as 
``technical alignment'' or ``policy focused''. Focusing on the technical side, sub-areas include \textit{Agent Foundations} (understanding the nature of intelligence using concepts like decision theory) \cite{Soares2017AgentFF}, \textit{Interpretability} (which can make alignment easier) \cite{defendersinterp, olah_2022}, \textit{Corrigibility} (making systems willing to be changed by their designers) \cite{soares2015corrigibility}, \textit{Prosaic Alignment} (working on alignment strategies of today's systems using today's tools; includes ideas like RL from Human Feedback \cite{christiano2023deep, ouyang2022training, bai2022training} or CAI \cite{bai2022constitutional} for aligning language models), \textit{Robustness} (ensuring alignment won't fail from out of distribution environments or from adversaries) \cite{DBLP:journals/corr/abs-2109-13916}. See \autoref{secextrarescources} for more comprehensive lists.

\subsection{Common Skepticisms}

Works like \cite{yampolskiy2022ai} and \cite{Russell2022ProvablyBA} survey skepticism on the need for focused AI safety work. Debates are not new; In 1950 \citet{turing1950mind} argued against skeptics of thinking machines. Categories from \cite{yampolskiy2022ai} include:

 \begin{enumerate}
     \item Priorities Objections:  \emph{HLMI problems are too futuristic} and so unimportant compared to other problems. This objection neglects evidence that timelines to HLMI are shortening, and that safety problems might take time to solve.
     \item Technical Objections: \emph{HLMI as unattainable}. Recent successes contradict this position.  
     Others mistakenly believe there will be trivial technical solutions to safety. 
     Others view AI safety an impossible problem to work on currently (a view we hope diminish with work like this).
     \item ``Biased Objections'': \emph{Safety considerations will slow progress or decrease funding}. This has echos of 20th-century biased research into tobacco or pollutants \cite{Oreskes2010MerchantsOD}. As AI advances, this is not sustainable.
 \end{enumerate}

Skepticism on advanced AI and existential risks is natural: these fears seem very \textit{weird} and can feel like science fiction. 
 Yet, significant evidence supports this weird future.

\subsection{SE Research's Unique Leverage}

SE researchers have an active role to play in long-term AI safety, due to the nature of software. Transformational and risky capabilities come from when machines can form and execute complex plans. It seems likely that reasoning on code will be a key skill which enables these capabilities. Highlighting this importance, a first multi-million user deployment of billion+ parameter neural networks came not for NLP or CV or RL, but for SE with Github Copilot.

Traditional research problems directed at making complex software behave as expected can be adapted to apply to AI software and also machine-written software.

\subsection{Additional References}\label{secextrarescources}

There are many survey papers on AI Safety \cite{Everitt2018AGISL, DBLP:journals/corr/abs-2109-13916, DBLP:journals/corr/abs-2002-05671, Sotala2014ResponsesTC, DWIVEDI2021101994, Russell2015ResearchPF, Critch2020AIRC}. In particular, \cite{Everitt2018AGISL} provides a nice general survey. \cite{DBLP:journals/corr/abs-2109-13916} has a useful emphasis on ML. \cite{Hendrycks2022XRiskAF} discusses assessing risks of AI work. \cite{Tambon2021HowTC, Dey2021MultilayeredRO} are SE journal articles on ML safety (but are less HLMI focused).

 AgiSafetyFundamentals.com collects a set of links and a course syllabus\footnote{\url{https://agisafetyfundamentals.com/resources}  and \href{https://www.agisafetyfundamentals.com/ai-alignment-curriculum}{../ai-alignment-curriculum}}. Berkeley CHAI similarly provides a bibliography\footnote{\url{https://humancompatible.ai/bibliography}}, ordered by topic and ``priority''.

Books such as \citet{bostrom2017superintelligence} collect key arguments in longer form. Other recent books include \cite{Russell2019HumanCA, Tegmark2017Life3B, Christian2021TheAP}.

\clearpage

\section{Subproblems}


We outline 4 conjectures about the future of software engineering, and for each discuss 3 example problems that might need to be solved to ensure safe progress. 
Addressing these challenges within the SE domain could help make other AI systems safer.

\subsection{Most Software Will Be Machine Written}

A nearer-term problem than HLMI is when a majority of source code is written by machines (at the abstraction level of current languages. Compiled and declarative high-level languages allow most code to be ``machine-written" as per 1950s abstraction levels).

Not long after deploying a generative AI autocomplete system, Google engineers reported in 2022 that approximately 3\% of new internal code was AI-written \cite{tabachnyk_nikolov_2022}. On the path to 50\% there are many safety-relevant problems.

\newcounter{ProbSECount}

\newcommand{\ProbSE}[1]{%
\refstepcounter{ProbSECount}
	\vspace{0.05in} \noindent \textbf{P\arabic{ProbSECount}:} ~\textit{#1} 
}

\ProbSE{How to reliably extract indicators of  uncertainty from code generators, and support auditing?}

In a world where most code is machine written, there needs to be ways to determine when the code might be misaligned with intent. A first step could accurately estimate the probability AI4SE outputs are correct. Work on calibrating ML probabilities includes \cite{Lakshminarayanan2016SimpleAS, Guo2017OnCO,   kadavath2022language}.
Careful attention is required: a claim like ``the system says the code is 99\% likely to be right, good enough'' implies likely failure within 100s of tries. More complex modeling of uncertainty should account for varying degrees of negative impact (a 10\% risk of over-counting the number of files in a directory is better than a 10\% risk of deleting all the files in a directory). It could also help to localize certain regions of code which are uncertain or under-specified (eg, if asked to generate a button, the system might choose a button color or size even if it is not provided in the prompt specification). A related approach is to generate code with ``holes" in uncertain regions \cite{Guo2021LearningTC}.
Learning about uncertainty might generalize to other generative AI systems with complex output (like in natural language (NL) or a robot action-space).

\ProbSE{How to create faithful summaries?}

Humans might not be able to review all lines of code; faithful summaries can help developers understand machine-written software. Summarizing complex software outputs might also have lessons for summarizing other AI created plans that are too complex for easy review. Prior safety-motivated work has explored summarizing longform NL like books \cite{Wu2021RecursivelySB}.

AI4SE has long studied automated summarization of source code \cite{Zhang2022ASO} or other artifacts like pull requests \cite{Liu2019AutomaticGO}. The studied datasets are traditionally human written artifacts, but a long-term view likely should put increasing emphasis on artifacts coming from generators. While the summarizer might fail, its failures should ideally be anti-correlated with generator failures. 

\ProbSE{Improve code provenance, accountability, and monitoring.}

In the current SE paradigm, version control systems like Git makes each line of code traceable to a human author. As we move to a world where most code is machine written, code committed by a  developer may not have been written by them;  
we thus need ways track what tool generated the code, under what conditions (\emph{e.g.,} prompting), and how it was audited. 
Commentary on this need has appeared previously \cite{10.1145/3582083}.

\subsection{Value of testing and verification increases.}

While testing is an important enabler of software reliability today, gaps in testing are
hopefully backstopped by having code written carefully by skilled and thoughtful humans. With improving code generators, we could expect that less human effort will be spent on implementation, and more on testing and verifying software. This conjecture is trackable with studies of developer time\footnote{eg, \cite{Meyer2019TodayWA} which estimated Microsoft developers spending 29\% of time spent on implementation focused activities vs 19\% on testing/reviewing/specification focused activities. If true we would expect time to shift away from implementation time.}.

\ProbSE{Identifying and helping writing tests for most critical code.}

Testing provides a formalized way to specify intent \cite{ticoder}. Thus, a potential way to improve safety is to work on ways of identifying the most important and uncertain regions of code, and guiding developers to formalize their intents as tests.

\ProbSE{Improved systems-level granularity and AI4SE safety claims.}

Modern operating systems and run-times are responsible for assuring that software only accesses resources (like memory, compute time, files, cameras, etc) that it is permitted. This improves trust in software written by others. As the ``others'' shift from humans to AI systems, the  need for
such OS assurances will likely increase. Safety might be improved if an AI system could make specific, verifiable, claims, accompanying generated code (\emph{e.g}, that it won't touch files, or have certain runtime bounds; Necula's proof-carrying code~\cite{necula1997proof} is a historical example). Those claims could then be enforced at a system level. 
Improved systems controls can likely improve software safety that is intermixed with AI written code, and be a useful proxy when thinking about controlling AI outputs across domains.

\ProbSE{Responsible automated vulnerability detection tools.}

There is interest in automated detection, prioritization, and repair of bugs and vulnerabilities \cite{Russell2018AutomatedVD, le2022survey}. This could have long-term benefits, but it's important to note that vulnerability detection is a dual-use technology (vulnerabilities can be maliciously exploited). The security research community is usually already well aware of dual-use risks, but there must be expanded awareness in the SE community that AI can make exploiting vulnerabilities more scalable. Yet, it seems there is a reasonable long-term safety argument towards the research as hardening defenses and fixing vulnerabilities helps stop rapid changes from the status quo (\emph{e.g.}, an AI system hacking into cloud computing providers to acquire vastly more compute) without time for intervention, or help prevent the hacking of aligned AGI \cite{hackagi}. 

\subsection{Everyone will be a software engineer.}

In the 1950s, computers were large and very expensive, and operating them was a specialized occupation. Today, with PCs, smartphones and embedded devices, everyone is a "computer operator". 

A natural conjecture is that on the path to HLMI, everyone will become a software engineer. 
Users will ``speak apps into existence'', an extreme form of WYSIWYG (What You See Is What You Get) or no-code/low-code editors that let users click websites, documents, or apps into existence.
This transformation is not just about autogenerated code, but everyone undertaking all parts of SE (problem specification, testing, deployment, etc).
Changes here are difficult to track, but estimates of low-code usage can be proxies \cite{Gartner2022}.

\ProbSE{Narrowing the PL$\leftrightarrow$NL gap}

There is a gap between the precision of a typical programming language and the ambiguity of natural language. Software creators of the future might not understand much about traditional programming languages. Challenges caused by this are apparent in existing low-code platforms with limited AI use \cite{rokis2022challenges}. One could imagine new programming languages or low-code paradigms that better mix together ambiguity and precision, and guide the user towards precision, particularly in areas with high risks of harms from intent ambiguity. While doing this, safety-conscious researchers could consider how their findings could generalize as systems scale to super-human capability or to other domains.

\ProbSE{Education of failure modes}

Effective techniques must be developed to educate non-technical users how automated SE tools can fail. This is a combination of the challenges of testing low-code \cite{Khorram2020ChallengesO}, explaining AI decisions \cite{Miller2017ExplanationIA, 10.1145/3387166}, and out-of-distribution detection \cite{Yang2021GeneralizedOD}.
A safety-conscious researcher can also think about how these education techniques can be used to teach about failure modes of broader AI. 

\ProbSE{Methods of dissuading overuse and modeling trust of systems}

It seems beneficial to understand when the use of  an AI4SE system should be discouraged; although this runs  counter to the natural goals of improving productivity, it is essential for a responsible system. Work like \cite{Cheng2022ItWW, Liao2022DesigningFR} has studied  trust of tools.

\vspace{-0.4em}
\subsection{AI written software will be used in the most critical parts of society.}

Software is an essential part of financial systems, health systems, and warfighting systems. As AI4SE becomes the norm, it will likely be used in writing software for the most critical parts of society. 

\ProbSE{Adapting SE techniques for ML reliability}

If complex ML-based systems are deployed in critical areas, there are likely opportunities to apply SE techniques for testing and understanding ML systems. Existing \cais in SE works are mostly in this SE4AI robustness area \cite{Gerasimou2020ImportanceDrivenDL, Baluta2020ScalableQV}. Other SE4AI work has seeked to understand how to deploy AI systems \cite{MartinezFernandez2021SoftwareEF,Gezici2022SystematicLR}. Improved robustness and understanding can make aligned HLMI more stable.




\ProbSE{Defining normative values of automated software tools?}

Systems like ChatGPT are helping raise awareness that designing systems require normative decisions \cite{weidinger2021ethical, Jakesch2023CoWritingWO, Ganguli2023TheCF}.  
There must be expanded understanding of similar failure modes and norms in AI4SE (for example, \cite{codex} explains how a code model might promote stereotypes of race or gender when writing algorithms or UIs). As machines author or edit software for the most important parts of society, the decisions become increasingly important. Additionally, proactive norms are needed in topics like self-modification of AI4SE tools or automated malware generation.

\ProbSE{Building regulation, policy, and safety-conscious culture.}

AI regulations are sometimes controversial \cite{Zhang2021EthicsAG, Michael2022WhatDN}. However, when considering the use of AI4SE systems in the most safety-critical systems, the need for common enforceable rules becomes increasingly clear. Natural market forces encourage some safety precautions, but likely are not enough to prompt adequate investment into avoiding catastrophes from highly capable AI systems, while also avoiding ``arms races'' \cite{dafoe2018ai}. 

AI poses many complicated problems that are going to take significant effort. Thus, wide community involvement and a culture of safety and preparedness is needed.

\subsection{Anti-problems for Safety}

Not all problems aid safety. We will discuss one example.

\newcounter{AntiProbSECount}

\newcommand{\AntiProbSE}[1]{%
\refstepcounter{AntiProbSECount}
	\vspace{0.1in} \noindent \textbf{Non-safety P\arabic{AntiProbSECount}:} ~\textit{#1} 
}

\AntiProbSE{Improving correctness metrics or user metrics is not necessarily net-beneficial to safety.}

A natural direction of the field is to drive forward correctness metrics of tasks like automated code generation. However, capability drives are not necessarily a safety objective. For example, improving the fraction of methods passing tests in 
the HumanEval \cite{codex} dataset
 (which 
measure ability to synthesize programs) is not a safety objective. Similarly, optimizing user metrics like the fraction of generations accepted is not a safety objective. On the surface, one might argue that making AIs generate more correct code improves safety, compared to the alternative of AI writing incorrect code. Yet, we claim this is a bit of a false dichotomy. There is an alternative with the current \emph{status quo} of human-authored code, which has many issues (high cost, inaccessible, error prone, etc), but gives more time to understand risks and safety problems. Pushing metrics like generation accuracy to their natural limit can lead to dangerously capable systems. Reaching the limit does not happen without incremental steps which on the surface seem harmless.

We are not advocating here the stronger view that progress on code tools or code generators should completely halt (there is huge upside to automating software development). Rather, we note that 
advances in correctness and capability, while important, don't necessarily lead to greater long-term safety.
Researchers, institutions, conferences, and the field should consider the portfolio of research being done. The proportion that is safety-focused should be brought in line with the amount of risk if the capabilities work on AI4SE succeeds and we reach the wild world where intelligent machines write most code in all parts of society.


We acknowledge that the distinction between this ``anti-problem'' and some of the other problems mentioned are not always ideally clear. Work on clearer boundaries and trade-offs is beneficial.

\section{Conclusions}

This set of conjectures and sub-problems is not intended to be comprehensive or definitive. Instead it is a set of example starting points. Solving these problems alone will not solve AI alignment (and defining what would is an open problem). There are also very real concerns of dual-use risks, and of ``safety-washing'' \cite{lesswrongSafetywashingLessWrong, lesswrongBewareSafetywashing} where safety impact gets mischaracterized. However, solving pragmatic problems can be stepping stones for more robust solutions while involving more communities.

We must work to break down taboos around seriously discussing HLMI and superintelligent AGI at SE venues (and all computing fields). We hope increasing numbers of SE researchers will reflect seriously on what the future of SE looks like, and then motivate their research questions not just on short-term progress, but also on its impact in a future of highly-capable AI.
\addtolength{\skip\footins}{1cm} 
\fancyfoot[L]{
\textcolor{white}{- \\ - \\}
Acknowledgements: DG's research work was partially supported by the NSF (CCF-1934568). We are thankful for colleagues and readers who gave feedback on draft ideas.}

\clearpage

\bibliographystyle{ACM-Reference-Format}
\bibliography{ais4sebib}

\fancyhf{}
\appendix


\end{document}

%% file: cite_vars.tex



\newcommand{\Numfwais}{44\xspace}
\newcommand{\FwaisTotalCitesDuplicates}{9198\xspace}
\newcommand{\FwaisTotalCitesDedup}{6565\xspace}

\newcommand{\PcSeTotal}{7744\xspace}
\newcommand{\PcSeCais}{4\xspace}
\newcommand{\PcAiTotal}{17906\xspace}
\newcommand{\PcAiCais}{87\xspace}
\newcommand{\PcMlTotal}{24731\xspace}
\newcommand{\PcMlCais}{225\xspace}
\newcommand{\PcCvTotal}{21192\xspace}
\newcommand{\PcCvCais}{17\xspace}
\newcommand{\PcNlpTotal}{15561\xspace}
\newcommand{\PcNlpCais}{40\xspace}
\newcommand{\PcPlTotal}{1550\xspace}
\newcommand{\PcPlCais}{0\xspace}
\newcommand{\PcArxivTotal}{199394\xspace}
\newcommand{\PcArxivCais}{665\xspace}
\newcommand{\PcUnknownTotal}{3055023\xspace}
\newcommand{\PcUnknownCais}{1405\xspace}
\newcommand{\PcOtherconfCais}{360\xspace}
\newcommand{\PcSejournalsTotal}{6082\xspace}
\newcommand{\PcSejournalsCais}{5\xspace}
\newcommand{\PcOtherjournalCais}{2742\xspace}
\newcommand{\OtherCais}{2810\xspace}

\newcommand{\SePercentAiOld}{4\%\xspace}
\newcommand{\SePercentAiNew}{33\%\xspace}

\newcommand{\fais}{\texttt{FSafe}\xspace}
\newcommand{\cais}{\texttt{CSafe}\xspace}

\newcommand{\Numcaiscsfield}{3484\xspace}